\documentstyle[12pt]{article}

\newcommand{\be}{\begin{equation}}
\newcommand{\ee}{\end{equation}}
\newcommand{\bea}{\begin{eqnarray}}
\newcommand{\eea}{\end{eqnarray}}

\makeatletter
\gdef\@pubnumber{\null}
\long\def\pubnumber#1{\gdef\@pubnumber{#1}}
\def\@makepub{\vbox to \z@{\hbox to \textwidth{\hfill
\llap{\parbox[t]{0.33\textwidth}{\raggedleft\@pubnumber}}}%
\vss}}
\def\@maketitle{\newpage
\@makepub \null
 \vskip 2em \begin{center}
 {\LARGE \@title \par} \vskip 1.5em {\large \lineskip .5em
\begin{tabular}[t]{c}\@author
 \end{tabular}\par}
 \vskip 1em {\large \@date} \end{center}
 \par
 \vskip 1.5em}

\title{Calculation of 130 GeV Mass for T-Quark}

\author{Frank D. (Tony) Smith, Jr.\thanks{\copyright
Frank D. (Tony) Smith, Jr.,
341 Blanton Road, Atlanta, Georgia 30342 USA.}\hspace{3pt},\
{\normalsize Department of Physics}\\
{\normalsize Georgia Institute of Technology}\\
{\normalsize Atlanta, Georgia 30332 USA}\\
{\normalsize e-mail address:  gt0109e at prism.gatech.edu}}

\pubnumber{THEP-93-2}

\date{January 1993}

\begin{document}

\maketitle

\begin{abstract}

	A 130 GeV mass is calculated for the t-quark, also called the
truth quark or the top quark.

	The 130 GeV T-quark mass is consistent with the analysis of
Dalitz and Goldstein [1] (131 GeV (-11,  +22 )) of a CDF T-quark candidate
event.  It is not consistent with the theoretical calculations
 of Dimopoulos, Hall, and Raby [2] that the T-quark mass range
should be 176 to 190 GeV.

	The calculation uses a model in which the 28-dimensional
adjoint representation of Spin(8) forms the gauge group of the model
with an 8-dimensional spacetime whose tangent vector space is
represented by the vector representation of Spin(8);
the first generation fermion particles and antiparticles are
represented by the two mirror image 8-dimensional half-spinor
representations of Spin(8).   The 8-dimensional spacetime of the model
is reduced to 4 dimensions.  Dimensional reduction gives the fermions
a 3-generation structure.
\end{abstract}

\section{Introduction}

		The model used in the theoretical calculations is based on the
Lie group Spin(8).

	The 28-dimensional adjoint representation of Spin(8) forms the
gauge group of the model with an 8-dimensional spacetime whose tangent
vector space is represented by the vector representation of Spin(8).

	The first generation fermion particles and antiparticles  are
represented by the two mirror image 8-dimensional half-spinor
representations of Spin(8), denoted by S8+ and S8-.

	The 8-dimensional spacetime of the model is reduced to 4 dimensions.

\vspace{12pt}

	After dimensional reduction: \\MacDowell-Mansouri gravity becomes
an effective nonrenormalizable theory;\\a natural Higgs scalar field
appears that gives mass to the weak bosons and Dirac fermions;\\and
there are three generations of fermions, being effectively represented by:

		octonions,

		pairs of octonions, and

		triples of octonions.

\section{3-Generation Structure of Fermions.}

	The model puts the 8 first generation fermion particles
in one 8-dimensional half-spinor representation S8+ of Spin(8),
with the 8 antiparticles being in the mirror image 8-dimensional
half-spinor representation S8- of Spin(8).
	Given a basis (1, i, j, k, ke, je, ke) for the octonions O ,
where 1 is the basis element for the real axis and, as to the seven
imaginaries, i, j, and k are just the three imaginary quaternions,
and e, ie, je, and ke are constructed from the four quaternionic
basis elements 1, i, j, and k by introducing an octonionic
imaginary e.

\vspace{12pt}

	The octonionic basis for S8+ corresponds to fermion
particles as follows:  1 is the electron neutrino;
i, j, k	 are the red, blue, and green up quarks;  e is the electron;
and ie, je, ke are the red, blue, and green down quarks.
The antiparticle correpsondence for S8- is similar.

\vspace{12pt}

	Consider the model from a lattice gauge theory point of view.
It looks like an 8-dim lattice of vertices connected by links.
The spinor fermions are assigned to the vertices of the lattice
spacetime.  The fermions go from place to place by moving from
an origin vertex to a destination vertex

(origin) *-----* (destination)
along a link that connects them.

\vspace{12pt}

The gauge bosons are assigned to the links, and are
represented as Lie group elements of the gauge group,
which then acts as a transport.
	Effectively,  dimensional reduction does not give a generation
structure to gauge bosons because the transport

*-----(g1)-----*-----(g2)-----* is the same as

*-----(g1 g2)-----*

where g1 and g2 are gauge Lie group
elements and g1 g2 is the Lie group product.

\vspace{12pt}

	Consider fermion particles (similar arguments apply
to antiparticles) represented by the 8 basis octonions
(1,i,j,k,e,ie,je,ke) of S8+.
For this discussion of how the three generations are formed,
ignore helicity and the distinction between the Weyl neutrino (1)
and the Dirac electron and quarks (i,j,k,e,ie,je,ke).

\vspace{12pt}

In 8-dim lattice F4 model, a fermion particle * going from
(origin) to (destination) can be represented by an octonion

(origin)*-----(destination), or in short

*----- , where * is an octonion.

\vspace{12pt}

	This notation uses *, o, and o' as notations for octonions
representing fermion particles at the vertices also denoted
by *, o, and o'.  Since the fermions live on vertices, the
abuse of notation (which is useful) should not be misleading.

\vspace{12pt}

What happens when the spacetime is reduced to 4 dimensions?

	The 8-dim E8 lattice,
    with octonionic (1,i,j,k,e,ie,je,ke) vertices,
goes to a 4-dim lattice  with quaternionic (1,i,j,k) vertices.

	The dimensional reduction is like a projection

(1,i,j,k,e,ie,je,ke)  -----  (1,i,j,k)

The subspace (1,i,j,k) is invariant and the subspace
(e,ie,je,ke)  is projected into  (1,i,j,k) .

\vspace{12pt}

Consider a given fermion particle going in 8-dim
(origin)*-----(destination)

\vspace{12pt}

THERE ARE 4 CASES:

CASE 1.  (origin) and (destination) are both in the (1,i,j,k)
subspace of (1,i,j,k,e,ie,je,ke).
Then the fermion:

(origin)*-----(destination) IS NOT CHANGED by
the dimensional reduction and is still represented by the
single octonion *.

\vspace{12pt}

THE CASE 1. FERMIONS ARE FIRST GENERATION FERMIONS.

\vspace{12pt}

CASE 2.  (origin) is in the (1,i,j,k) subspace of
(1,i,j,k,e,ie,je,ke) BUT (destination) has components in the
(e,ie,je,ke) subspace of (1,i,j,k,e,ie,je,ke).
Then the fermion:

(origin)*-----(destination) IS CHANGED
because dimensional reduction takes the (destination)o vertex o
in (1,i,j,k,e,ie,je,ke) into its image in (1,i,j,k) under the
dimensional reduction map, denoted as (reduced destination).
    The result is
fermion:

(origin)*-----(destination)o-----(reduced destination).

    After dimensional reduction, there is a new intermediate
vertex o on which another octonion fermion can live.

    THEREFORE, IT TAKES TWO (2) OCTONIONS TO REPRESENT SUCH
A CASE 2. FERMION AFTER DIMENSIONAL REDUCTION.

    IT IS A SECOND GENERATION FERMION, REPRESENTED BY
A PAIR (*,o) OF OCTONIONS.

\vspace{12pt}

CASE 3.  (origin) has components in the (e,ie,je,ke) subspace
     of (1,i,j,k,e,ie,je,ke)
BUT (destination) is in the (1,i,j,k) subspace
     of (1,i,j,k,e,ie,je,ke).
Then the fermion:

(origin)*-----(destination) IS CHANGED
because dimensional reduction takes the (origin)* vertex *
in (1,i,j,k,e,ie,je,ke) into its image in (1,i,j,k) under the
dimensional reduction map, denoted as (reduced origin)o,
before the fermion goes to its (destination) in (1,i,j,k).
    The result is
fermion:

(origin)*-----(reduced origin)o-----(destination).

    After dimensional reduction, there is a new intermediate
vertex o on which another octonion fermion can live.

     THEREFORE, IT TAKES TWO (2) OCTONIONS TO REPRESENT SUCH
A CASE 3. FERMION AFTER DIMENSIONAL REDUCTION.

	IT IS ALSO A SECOND GENERATION FERMION, REPRESENTED BY
A PAIR (*,o) OF OCTONIONS.

The only remaining possibility is

\vspace{12pt}

CASE 4.  (origin) has components in the (e,ie,je,ke) subspace
    of (1,i,j,k,e,ie,je,ke)
AND (destination) has components in the (e,ie,je,ke) subspace
    of (1,i,j,k,e,ie,je,ke).
Then the fermion:

(origin)*-----(destination) IS CHANGED
because dimensional reduction takes the (origin)* vertex *
in (1,i,j,k,e,ie,je,ke) into its image in (1,i,j,k) under the
dimensional reduction map, denoted as (reduced origin)o,
before the fermion goes to its (destination) in (1,i,j,k,e,ie,je,ke),
which is denoted by (destination)o'.

	THEN THE DIMENSIONAL REDUCTION MAP takes
the (destination)o' vertex o' in (1,i,j,k,e,ie,je,ke) into its image in
(1,i,j,k) under the dimensional reduction map,
denoted as (reduced destination).
    The result is fermion:

(origin)*----(reduced origin)o----(destination)o'----(reduced destination).

    After dimensional reduction, there is are two (2) new intermediate
vertices o and o' on which two more octonion fermions can live.

    THEREFORE, IT TAKES THREE (3) OCTONIONS TO REPRESENT SUCH
A CASE 4. FERMION AFTER DIMENSIONAL REDUCTION.

    IT IS A THIRD GENERATION FERMION, REPRESENTED BY
A TRIPLE (*,o,o') OF OCTONIONS.

\vspace{12pt}

SINCE THERE ARE NO MORE CASES, THERE ARE ONLY 3
GENERATIONS.

\vspace{12pt}

Gauge bosons do not get a 3-generation structure because
the corresponding pair or triple of links (g1 , g2)  or
(g1 , g2 , g3)  can be reduced to a single gauge boson by the
Lie group product g1 g2  or g1 g2 g3.
(On a finite lattice, gauge Lie group elements,
not infinitesimal Lie algebra elements, live on the links.)

The important difference here between adjoint rep gauge bosons and
spinor rep fermions is that the adjoint rep gauge bosons
inherit the gauge Lie group product,
and the spinor rep fermions have no such product.

\section{First-Generation Quark Consitituent Masses.}
	In the model, the Weyl fermion neutrino has at tree level
only the left-handed state, whereas the Dirac fermion electron
and quarks can have both left-handed and right-handed states,
so that the total number of states corresponding to each of the
half-spinor Spin(8) representations Spin(8) is 15.

	Neutrinos are massless at tree level in all generations.

\vspace{12pt}

	In the model, the first generation fermions correspond to
octonions O, while second generation fermions correspond to pairs of
octonions OO and third generation fermions correspond to triples of
octonions OOO.

\vspace{12pt}

	To calculate the fermion masses in the model, the volume of a
compact manifold representing the spinor fermions S8+is used.
It is the parallelizable manifold $S^7\times RP^1$ .

	Also, since gravitation is coupled to mass, the infinitesimal
generators of the MacDowell-Mansouri gravitation group, Spin(5),
are used in the fermion mass calculations.

	The calculated quark masses are constituent masses, not
current masses.

\vspace{12pt}

	In the model, fermion masses are calculated as a product
of four factors:

 	$V(Q)\times N(Graviton)\times N(octonion)\times Sym$

\vspace{12pt}

	V(Q) is the volume of the part of the half-spinor fermion
particle manifold $S^7\times RP^1$ that is related to the fermion
particle by photon, weak boson, and gluon interactions.

\vspace{12pt}

	N(Graviton) is the number of types of Spin(5) graviton related to
the fermion.  The 10 gravitons correspond to the 10 infinitesimal
generators of Spin(5) = Sp(2).  2 of them are in the Cartan subalgebra.
6 of them carry color charge, and may therefore be considered as
corresponding to quarks.  The remaining 2 carry no color charge, but
may carry electric charge and so may be considered as corresponding
to electrons.

	One graviton takes the electron into itself, and the other can
only take the first-generation electron into the massless electron
neutrino.   Therefore only one graviton should correspond to the mass
of the first-generation electron.

	The graviton number ratio of the down quark to the
first-generation electron is therefore 6/1 = 6.

\vspace{12pt}

	N(octonion) is an octonion number factor relating up-type quark
masses to down-type quark masses in each generation.

\vspace{12pt}

	Sym is an internal symmetry factor, relating  2nd and 3rd
generation massive leptons to first generation fermions.  It is not used
in first-generation calculations.

\vspace{12pt}

	The ratio of the down quark constituent mass to the electron mass
is then calculated as follows:
	Consider the electron, e.  By photon, weak boson, and gluon
interactions, e can only be taken into 1, the massless neutrino.
The electron and neutrino, or their antiparticles, cannot be combined to
produce any of the massive up or down quarks.  The neutrino, being
massless, does not add anything to the mass formula for the electron.
Since the electron cannot be related to any other massive Dirac
fermion, its volume V(Q) is taken to be 1.

\vspace{12pt}

	Next consider a red down quark ie.  By gluon interactions, ie can
be taken into je and ke, the blue and green down quarks.  By weak
boson interactions, it can be taken into i, j, and k, the red, blue, and
green up quarks.  Given the up and down quarks, pions can be formed
from quark-antiquark pairs, and the pions can decay to produce
electrons and neutrinos.  Therefore the red down quark (similarly, any
down quark) is related to any part of $S^7\times RP^1$, the compact
manifold corresponding to (1, i, j, k, e, ie, je, ke), and therefore a
down quark should have a spinor manifold volume factor of the volume
of $S^7\times RP^1$.
	The ratio of the down quark spinor manifold volume factor to
the electron spinor manifold volume factor is just
V($S^7\times RP^1$)/1 = $\pi ^5$/3 .

	Since the first generation graviton factor is 6,

md/me = 6V($S^7\times RP^1$) = $2\pi ^5$ = 612.03937.

\vspace{12pt}

	As the up quarks correspond to i, j, and k, which are isomorphic
to ie, je, and ke of the down quarks, the up quarks and down quarks
have the same constituent mass mu = md.

	Antiparticles have the same mass as the corresponding
particles.

\vspace{12pt}

	Since the model only gives ratios of massses, the mass scale is
fixed by assuming that the electron mass me = 0.5110 MeV.  Then, the
constituent mass of the down quark md = 312.75 MeV, and the
constituent mass for the up quark mu = 312.75 MeV.

	As the proton mass is taken to be the sum of the constituent
masses of its constituent quarks, m(proton) = mu + mu + md = 938.25 MeV,
the model calculation is close to the experimental value of 938.27 MeV.

\section{T-Quark Mass Calculation.}

	The third generation fermion particles correspond to triples of
octonions.  There are $8^3$ = 512 such triples.
	The triple (1,1,1) corresponds to the tau-neutrino.
	The other 7 triples involving only 1 and e correspond to the
tauon: (e,e,e), (e,e,1), (e,1,e), (1,e,e), (1,1,e), (1,e,1), and (e,1,1).

	The symmetry of the 7 tauon triples is the same as the
symmetry of the 3 down quarks, the 3 up quarks, and the electron,
so the tauon mass should be the same as the sum of the masses of
the first generation massive fermion particles.

	Therefore the tauon mass 1.87704 GeV.

	Note that all triples corresponding to the tau and the tau-neutrino
are colorless.

\vspace{12pt}

	The beauty quark corresponds to 21 triples.  They are triples of
the form (1,1,ie), (1,ie,1), (ie,1,1), (ie,ie,1), (ie,1,ie), (1,ie,ie), and
(ie,ie,ie), and the similar triples for 1 and je and for 1 and ke.

	Note particularly that triples of the type (1,ie,je), (ie,je,ke), etc.,
do not correspond to the beauty quark, but to the truth quark.

\vspace{12pt}

	The red beauty quark is defined as the seven triples (1,1,ie), (1,ie,1),
(ie,1,1), (ie,ie,1), (ie,1,ie), (1,ie,ie), and (ie,ie,ie), because ie is the
red down quark.  The seven triples of the red beauty quark correspond
to the seven triples of the tauon, except that the beauty quark interacts
with 6 Spin(5) gravitons while the tauon interacts with only two.
The beauty quark constituent mass should be the tauon mass times the
third generation graviton factor 6/2 = 3, so the B-quark mass is
5.63111 GeV.

	The blue beauty quarks correspond to the seven triples involving
je, and the green beauty quarks correspond to the seven triples involving ke.

\vspace{12pt}

	The truth quark corresponds to the remaining 483 triples, so the
constituent mass of the red truth quark is 161/7 = 23 times the
red beauty quark mass, and the red T-quark mass is 129.5155 GeV.

	The blue and green truth quarks are defined similarly.

The tree level T-quark constituent mass rounds off to 130 GeV.

\section{Gauge Bosons and Further Results}

	Dimensional reduction also acts on the gauge bosons, giving
an effective SU(3) x SU(2)L x U(1) standard model gauge group
plus an effective Spin(5) MacDowell-Mansouri gravity.

\vspace{12pt}

	The physically realistic way to decompose the 28 infinitesimal
generators of Spin(8) after dimensional reduction is to group them
according to Weyl group symmetry into groups of 10, 6, 8, and 4 members.

	Then the group of 10 becomes Spin(5) with base manifold $S^4$,
the group of 6 becomes Spin(4) with base manifold $S^2\times S^2$,
the group of 8 becomes SU(3) with base manifold $CP^2$, and the
group of 4 becomes $U(1)^4$ with base manifold $T^4$.
The Weyl group of Spin(8) is the semidirect product of the
Weyl groups of the groups into which Spin(8) is decomposed.
Each group is then considered to be independent, with the effect
that the Spin(5) gives MacDowell-Mansouri gravity,

the SU(3) is the color force SU(3),

the Spin(4) has two copies of SU(2), one of which becomes the
effective weak force SU(2)L and the other of which is integrated
over the 4 "lost" dimensions to give an effective Higgs scalar field,
and

the 4 copies of U(1) become the 4 covariant components of the
electromagnetic photon.

\vspace{12pt}

	Second-generation fermion masses (constituent
masses for quarks), force strengths, the Weinberg angle,
and Kobayashi-Maskawa parameters can also be calculated
using the model, with the following results [3]:

\vspace{12pt}

$m_\mu$ = 104.8 MeV

$m_{\mu -neutrino}$ = 0

$m_s$ = ...625 MeV

$m_c$ = 2.09 GeV

\vspace{12pt}

	U(1) electromagnetism:
$\alpha _E={1 \over {137.03608}}$

\vspace{12pt}

   SU(2) weak force:
 $G_F = G_Wm_{proton}^2 = 1.02\times 10^{-5}$

\vspace{12pt}

	SU(3) color force:

$\alpha_{C}$ = 0.629 at 0.24 GeV

$\alpha_{C}$ = 0.168 at 5.3 GeV

$\alpha_{C}$ = 0.122 at 34 GeV

$\alpha_{C}$ = 0.106 at 91 GeV

\vspace{12pt}

	Higgs scalar mass = 260.8 GeV

\vspace{12pt}

$m_{W+}= m_{W-}= 80.9 GeV$

$m_{Z}=92.4 GeV$

\vspace{12pt}

Weinberg angle:
$Sin^2\theta _W=0.233$

\vspace{12pt}

	The  Kobayashi-Maskawa Parameters are:

	phase angle:
$e={\pi  \over 2}$

$V_{ud}$ = 0.975

$V_{us}$ = 0.222

$V_{ub}$ = -0.00461 i

$V_{cd}$ = -0.222 -0.000190 i

$V_{cs}$ = 0.974 -0.0000434 i

$V_{cb}$ =  0.0423

$V_{td}$ = 0.00941 -0.00449 i

$V_{ts}$ =  -0.0413 -0.00102 i

$V_{tb}$ = 0.999

	The same K-M mixing angles apply to both leptons and quarks,
but are only effective for leptons if neutrinos have nonzero mass.

	Beyond tree level, neutrinos can get mass by radiative processes
related to the Planck mass [3]:

$m_{e-neutrino}=2.2\times 10^{-6} eV$

$m_{\mu -neutrino}=4.5\times 10^{-4} eV$

$m_{\tau -neutrino}=8.1\times 10^{-3} eV$

	Therefore, the model has a natural MSW mechanism that
may solve the solar neutrino problem.

\section{Chronology of T-quark Mass Calculations}

\vspace{12pt}

At the request of others who have done theoretical
calculations, I am adding this section in January 1993:

\vspace{12pt}

I do not represent that this section is a complete
history of calculations of the T-quark mass.  It is
about the chronology of some theoretical T-quark mass
calculations of which I am now aware.
It includes only purely theoretical calculations
giving a result of about 130 GeV,
and does not include calculations of bounds on the
T-quark mass resulting from applying the standard model
(or other models) to experimental results such as
B-mixing, Z-width, etc.

\vspace{12pt}

{\bf 1982}:

Harvey, Reiss, and Ramond write
Mass Relations and Neutrino Oscillations in an SO(10) Model
(Nuc. Phys. {\bf B}{\bf 199} (1982) 223-268) (revised 3 Feb 82).

Eq. 3.33, tan $\alpha$ = $\sqrt{m_c \over m_t}$,
where tan $\alpha$ = $V_{cb}$,
relates the T-quark mass $m_t$ to the K-M parameter $V_{cb}$
and the C-quark mass $m_c$.

As the paper states (p. 237):

"Unfortunately, it does not
predict $m_t$ except through $\tau_B$ ... ."

\vspace{12pt}

Inoue, Kakuto, Komatsu, and Takeshita write
Aspects of Grand Unified Models with
Softly Broken Suypersymmetry
(Prog. Theor. Phys. {\bf 68} (1982) 927) (received 10 May 82)

They relate supersymmetry to electro-weak symmetry
breaking by radiative corrections and renormalization
group equations, and find that the renormalization
group equations have a fixed point.

The fixed point is related to a T-quark mass of
about 125 GeV, as was explicitly discussed in 1983 by
Alvarez-Gaume, Polchinski, and Wise.

\vspace{12pt}

{\bf 1983}:

Alvaerez-Gaume, Polchinski, and Wise write
Minimal Low-Energy Supergravity

(Nuc. Phys. {\bf B}{\bf 221} (1983) 495-523) (received 8 Feb 83).

Their calculations show that,
for electro-weak symmetry breaking to occur,
the T-quark mass must be from 100 GeV to 195 GeV.

Moreover, they also note (p. 511) that the renormalization
group equation

"... tends to attract the top quark mass
towards a fixed point of about 125 GeV."

\vspace{12pt}

Work similar to that of Alvarez-Gaume, Polchinski, and
Wise was done by Ibanez and Lopez in
N=1 Supergravity, the Weak Scale and the Low-Energy
Spectrum

(Nuc. Phys. {\bf B}{\bf 233} (1984) 511-544) (received 8 Aug 83).

\vspace{12pt}

As far as I know, the works
of Inoue, Kakuto, Komatsu, and Takeshita;
of Alvarez-Gaume, Polchinski, and Wise; and
of Ibanez and Lopez
are the first purely theoretical calculations
of the T-quark mass to be about 130 GeV.

\vspace{12pt}

{\bf 1984}:

I wrote Particle Masses, Force Constants, and Spin(8)

(Int. J. Theor. Phys. {\bf 24} (1985) 155-174) (received 27 Feb 84).

Using a model similar, but not identical, to the model
I am now using, I calculated the T-quark mass to
be 129.5 GeV.  (The models are similar with respect to
the T-quark mass.)

\vspace{12pt}

Nature ({\bf 310} (12 July 84) 97) article about CERN discovering
the T-quark at 40 GeV.

\vspace{12pt}

At the 31 Oct- 3  Nov 84 APS DPF Santa Fe meeting
I gave a 10-minute talk about my theoretical work
on the T-quark mass, and interpreted the CERN
experimental results as being consistent with
T-quark mass of 120 to 160 GeV, rather than 40 GeV.

\vspace{12pt}

{\bf 1986}:

Mohapatra, in his book
{\em Unification and Supersymmetry,
The Frontiers of Quark-Lepton Physics}
(Springer-Verlag 1986), in Section 15.3 on
Electro-Weak Symmetry Breaking and Supergravity,
discussed the work of Alvarez-Gaume, Polchinski, and Wise,
and stated (pp. 287-288 (323-324 in 1992 second edition)):

"It is interesting that $m_t$ lies in the range
100 GeV $\leq m_t \leq$ 190 GeV.  The recent discovery of
the t-quark in the mass range of 40-60 GeV therefore
rules out the simple-minded analysis carried out here."

\vspace{12pt}

{\bf 1991}:

Arason, Castano, Kesthelyi, Mikaelian, Piard, Ramond,
and Wright write

Top-Quark and Higgs-Boson Mass Bounds from a Numerical
Study of Sypersymetric Grand Unified Theories
(Phys. Rev. Lett. {\bf 67} (1991) 2933-2936) (received 7 Aug 91).

They calculate T-quark and Higgs masses
for SUSY scale of 1 TeV to be:

if $m_b$ = 4.6 GeV then $162 \leq m_t \leq 176$ GeV and
$106 \leq m_H \leq 111$ GeV; and

if $m_b$ = 5.0 GeV then $116 \leq m_t \leq 147$ GeV and
$93 \leq m_H \leq 101$ GeV.

\vspace{12pt}

{\bf 1992}:

I submitted a preprint to SLAC (10 July 92)

T PRINT-92-0226 [GEORGIA-TECH]  and

ET PRINT-92-0227 [GEORGIA-TECH] in SLAC index

and to CERN (week beginning 22 Sep 92)

PRE 33611 in CERN index.

This preprint is an extensive presentation of
my current work,

including T-quark mass = 130 GeV.

\vspace{12pt}

Arason, Castano, Piard, and Ramond give a renormalization
group analysis of mixing angles and masses
including the T-quark mass
in Phys. Rev. {\bf D} {\bf 47} (1993) 232-240.

\vspace{12pt}

Supersymmetric theories have been constrained by
CDF to have

squarks $\geq$ 126 GeV and gluinos $\geq$ 141 GeV

(Search for Squarks and Gluinos from p bar-p Collisions
at $\sqrt{s }$ = 1.8 TeV, Phys. Rev. Lett. {\bf 69} (1992) 3439-3443)
(received 17 Aug 92).

\vspace{12pt}

{\bf 1993}:
I submitted the replaced version of this paper to
hep-ph@xxx.lanl.gov as hep-ph/9301210 and to
clf-alg@stars.sfsu.edu bulletin boards.  (Jan 93)

\vspace{12pt}
\vspace{12pt}

	Further details of my work [3] are available as
a paper preprint or

as a Mathematica 2.1 notebook on a Macintosh HD disk.

\end{document}